\documentclass[seceq]{ptptex}

\usepackage{graphicx}

\title{
Emergent Einstein Universe under Deconstruction
}

\subtitle{Self-consistent geometry induced in theory space} 

\author{Nahomi \textsc{Kan}$^a,$%
\footnote{E-mail: kan@yamaguchi-jc.ac.jp} and
Kiyoshi \textsc{Shiraishi}$^b,$%
\footnote{E-mail: shiraish@yamaguchi-u.ac.jp}}
\inst{$^a$Yamaguchi Junior College, 
Hofu
, Yamaguchi 747-1232, Japan\\
$^b$Yamaguchi University, 
Yamaguchi
, Yamaguchi 753-8512, Japan
}

\abst{
We study self-consistent static solutions for an Einstein universe in
 a graph-based induced gravity. The one-loop quantum action is computed at finite
temperature. In particular, we demonstrate specific results for the models based on
cycle graphs. }

\begin{document}
\maketitle

\section{Introduction}
Yet we do not know any consistent quantum theory of gravity.
While the success of the combined 
model of three fundamental interactions
described by gauge theories has been achieved for many years ago, only gravitational
interaction is left unbound today. 
An eccentric
idea that the gravitational interaction can be
induced by quantum fluctuations of matter fields brings about almost forty years ago.
There are many versions of such `induced gravity' (or `emergent gravity').%
\footnote{For a review, see Ref \citen{IG}.}
In most cases, explicit values for fundamental constants cannot
be calculated because of the cutoff dependence, even if higher-loop contributions are
neglected.

In our previous paper\cite{KSPTP}, it is shown that calculable models of
induced gravity can be constructed based on the knowledge of spectral graph
theory.\cite{Mohar}
The Newton constant and the cosmological constant can be calculated
at one-loop level in flat-space limit.
Based on this work,
we consider self-consistent equations for a `graph
theory space' in the present paper. We demonstrate specific results for models based on
cycle graphs, which represent the original moose diagram in dimensional deconstruction.
\cite{Deconstruction}

The present paper is organized as follows.
In \S~2, we will start with reviewing background matters, {\it i.e.}, the basic ideas
behind the present work and techniques used in the following sections.  
Our previous work is briefly summarized in \S~3, for convenience. 
In \S~4, it is shown that the heat kernel method is used to evaluate the
self-consistent solutions in  specific models. The results for the models are
shown in \S~5. We give summary with some outlook in the last section.

We use the metric signature $(-+++)$ throughout this paper.

\section{The background matters of our work}

\subsection{Induced Gravity}
{\it Induced gravity} or {\it emergent gravity} has been studied by many
authors.\cite{IG} The idea of induced gravity is,
``Gravity emerges from the quantum effect of matter fields.''
In the terms of the heat kernel method\cite{hk}, the one-loop effective action can
schematically be expressed as the form: 
\begin{equation}
-\frac{1}{2}\int\frac{dt}{t}\sum_i {\rm str}\exp\left[-(-\Box+M^2_i)t\right]\,.
\label{fhk}
\end{equation}
Here $\Box$ is the d'Alembertian and $M_i$ is the mass of the $i$-th field.
They appear in the wave equation of the field.%
\footnote{Precisely speaking, the ``{d'Alembertian}'' comes from the second
functional derivative of the action with respect to the field. Thus the
``{d'Alembertian}'' for spinor and vector fields include the contribution from
curvature of spacetime in addition to the differential operator.}  The s-trace
(str) in (\ref{fhk}) is considered to be the trace including a sign dependent on the
statistics of the field. In curved four-dimensional spacetime, the s-trace part
including the
\begin{equation}
{\rm str}\,\exp\left[-(-\Box)t\right]=\frac{\sqrt{|\det g|}}{(4\pi)^{2}}\,t^{-2}
(a_0+a_1 t+\cdots)\,,
\end{equation}
where 
$g$ is the determinant of the metric and the
coefficients depend on the background fields.
In four-dimensional spacetime, the values of coefficients are
$a_0=1$ and
$a_1=R/6$ for minimal scalar fields,
$a_0=-4$ and $a_1=R/3$ for Dirac fields, 
$a_0=3$ and $a_1=-R/2$ for massive vector fields,
where $R$ is the scalar curvature of the spacetime.
The coefficient $a_1$ 
leads to the Einstein-Hilbert term, while $a_0$ yields the cosmological term.

In Kaluza-Klein (KK) theories,
inducing Einstein-Hilbert term was also studied.\cite{Toms} 

\subsection{Dimensional Deconstruction}
The concept of {\it dimensional deconstruction} (DD)\cite{Deconstruction} is equivalent
to  considering a higher-dimensional theory with discretized extra dimensions
at a low energy scale.

For example, let us consider a $SU(m)^N$ gauge theory. The lagrangian density for
vector fields is
\begin{equation}
\mathcal{L}=-\frac{1}{2e^2}{\rm Tr}\sum_{k=1}^N F_{\mu\nu
k}^2-\frac{2}{e^2}{\rm Tr}\sum_{k=1}^N
\left(D_\mu U_{k,k+1}\right)^2\,,
\end{equation}  
where $F_{k}^{\mu\nu}=\left[\partial^\mu-iA^\mu_k, \partial^\nu-iA^\nu_k\right]$ 
is the
field strength of $SU(m)_k$
and $\mu , \nu = 0,1,2,3$,
while $e$ is a common gauge coupling constant. 
We should read $A_{N+k}^\mu = A_{k}^\mu ,~etc$.  
$U_{k,k+1}$, called `a link field', is transformed as
\begin{equation}
U_{k,k+1} \to L_k U_{k,k+1} L_{k+1}^\dagger\qquad (L_k \in SU(m)_k)\,,
\label{transformation-U}
\end{equation}
under $SU(m)_k$.
The covariant derivative is defined as
\begin{equation}
D^\mu U_{k,k+1} \equiv \partial^\mu U_{k,k+1} - i A^\mu_k U_{k,k+1} + i U_{k,k+1}
A^\mu_{k+1}\,.
\end{equation}

A polygon with $N$ edges, called a `moose' diagram, is used to describe this theory,
in other words, to characterize the transformation (\ref{transformation-U}).  This
diagram is  consisting of `sites' and `links', which are usually represented by open
circles and single directed lines attached to these circles, respectively.
A gauge transformation is assigned to each site and
four-dimensional fields are assigned to
sites and links, in general.
The geometrical figure built up from sites and links (and also sometimes 
faces) is
sometimes called `theory space'. \cite{Deconstruction}

Turning back to our example,
we assume that the absolute value of each link field
$|U_{k,k+1}|$ takes the same scale, $f$.
Then $U_{k,k+1}$ is expressed as
\begin{equation}
 U_{k,k+1} = f \exp (i {\chi} _{k}/f) \,.
\end{equation}
We then find that the kinetic terms of $U_{k,k+1}$ go over to a mass-matrix for the
gauge fields. Since the gauge boson $(mass)^2$ matrix for $N=5$ turns out to be
\begin{equation}
f^2
\begin{pmatrix}
2      &    -1  &      0 &      0 &    -1  \\
-1     &     2  &     -1 &      0 &     0   \\
0      &    -1  &      2 &     -1 &     0   \\
0      &     0  &     -1 &     2  &    -1   \\
-1     &     0  &      0 &    -1  &     2  
\end{pmatrix} \,,
\label{gauge-boson_mass-matrix}
\end{equation}
we obtain the gauge boson mass spectrum 
\begin{equation}
M_{p}^{2} = 4f^2 \sin ^2 \left(\frac{\pi p}{N} \right)\qquad
(p \in {\bf Z}) \,,
\label{gauge-boson_mass-spectrum}
\end{equation}
by diagonalizing (\ref{gauge-boson_mass-matrix}).

For $|p| \ll N$, the mass spectrum becomes
\begin{equation}
M_p \simeq  \frac{2\pi |p|f}{N}\,.
\label{KK_mass-spectrum}
\end{equation}
If we define 
\begin{equation}
b= \frac{1}{f} \,, \quad \ell=Nb \,,
\label{aR}
\end{equation}
then the masses (\ref{KK_mass-spectrum}) can be written as
\begin{equation}
M_p \simeq \frac{2\pi |p|}{\ell} \,.
\end{equation}
This is precisely the KK spectrum for a five-dimensional gauge boson
compactified on a circle of circumference $\ell$.
In the limit of a large number of sites, 
$N \to \infty$,
DD leads to a five-dimensional theory,
where the extra space is a circle.


\subsection{Spectral Graph Theory}
Sites can be, in general, more complicatedly connected by links.
The theory space does not necessarily have a continuum limit 
and the diagram would have complicated connections or links.
Such a diagram is called as a {\it graph}. 
The $N$-sided polygon is identified as a simple graph, a {\it cycle graph} $C_N$. 

We identify the moose diagram of the theory space as a graph 
consisting of vertices and edges,
which correspond to sites and links, respectively.
Therefore,
DD can be generalized to {\it field theory on a graph}.\cite{KSJMP}

A graph $G$ consists of a {\it vertex} set 
and  an {\it edge} set 
where an edge is a pair of distinct vertices of $G$.
The {\it degree} of a vertex $v$, denoted by $deg(v)$, 
is the number of edges incident with $v$.

We will argue about the orientation of an edge. The graph with directed edges is dubbed
as a {\it directed graph}. An oriented edge $e=[u,v]$ 
connects the {\it
origin} $u=o(e)$ and the {\it terminus} $v=t(e)$.

{\it Spectral graph theory} is the mathematical study of a graph through
investigating various properties on eigenvalues, and eigenvectors of matrices
associated with it.\cite{Mohar}
Now we introduce various matrices which are naturally associated with a
graph.\cite{KSJMP,Mohar}
   
The {\it incidence matrix} $E(G)$ is defined as 
\begin{equation}
(E)_{ve} = \begin{cases}
            1 &  \text{if $v=o(e)$ }           \\ 
            -1 &  \text{if $v=t(e)$ }          \\ 
            0 &  \text{otherwise}
         \end{cases} \, .
\end{equation}
The {\it adjacency matrix} $A(G)$ is defined as \\
\begin{equation}
(A)_{vv^{\prime}} = \begin{cases}
            1 &  \text{if $v$ is adjacent to $v^{\prime}$ }           \\ 
            0 &  \text{otherwise}
         \end{cases} ~ .
\end{equation}
The {\it degree matrix} $D(G)$ is defined as \\
\begin{equation}
(D)_{vv^{\prime}} = \begin{cases}
           deg(v) &  \text{if $v = v^{\prime}$ }           \\ 
            ~~ 0   &  \text{otherwise}
         \end{cases} ~ .
\end{equation}
Note that ${\rm Tr}\,A=0$ and ${\rm Tr}\,A^2={\rm Tr}\,D$.

The {\it graph Laplacian} (or {\it combinatorial Laplacian}) 
$\Delta (G)$ is defined by \\
\begin{equation}
(\Delta)_{vv^{\prime}} 
      = (D-A)_{vv^{\prime}}
      = \begin{cases}
            deg(v) &  \text{if $v = v^{\prime}$ }           \\ 
             -1    &  \text{if $v$ is adjacent to $v^{\prime}$}        \\
            ~~ 0   &  \text{otherwise}
         \end{cases} ~ .
\end{equation} 
The most important fact is                       
\begin{equation}
\Delta= EE^T\,,
\label{EE}
\end{equation} 
where $E^T$ is the transposed matrix of $E$.
The Laplacian matrix is symmetric, so its eigenvalues are non-negative.
Note also that ${\rm Tr}\,\Delta={\rm Tr}\,D$ and
${\rm Tr}\,\Delta^2={\rm Tr}\,D^2+{\rm Tr}\,D$.

\begin{figure}[h]
\centering
\includegraphics[height=5cm]
{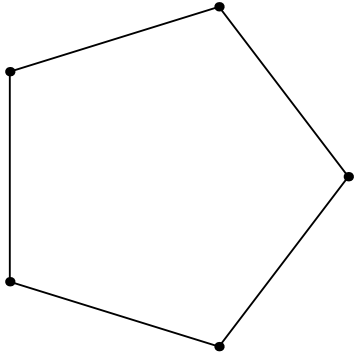}
\caption{%
The graph $C_{5}$.
}
\label{fig1}
\end{figure}

For a concrete example, we consider a {cycle graph},
which is equivalent to a popular moose diagram in DD, usually expressed as a polygon.
The cycle graph with $p$ vertices is denoted by $C_p$.
For $C_5$ (Fig \ref{fig1}).

The incidence matrix for $C_5$ can be written, if the orientation of edges are
`cyclic', as
\begin{equation}
E(C_5)=
\begin{pmatrix}
1  & 0  & 0  & 0  &-1 \\
-1 & 1  & 0  & 0  &0 \\
0  & -1 & 1  & 0  &0 \\
0  & 0  & -1 & 1  &0 \\
0  & 0  & 0  & -1 & 1
\end{pmatrix} \, .
\end{equation}

The adjacency matrix for $C_5$ takes the form
\begin{equation}
A(C_5)=
\begin{pmatrix}
0 & 1 & 0 & 0 & 1 \\
1 & 0 & 1 & 0 & 0 \\
0 & 1 & 0 & 1 & 0 \\
0 & 0 & 1 & 0 & 1 \\
1 & 0 & 0 & 1 & 0
\end{pmatrix} \, .
\end{equation}

The degree matrix for $C_5$ takes the form
\begin{equation}
D(C_5)=
\begin{pmatrix}
2 & 0 & 0 & 0 & 0 \\
0 & 2 & 0 & 0 & 0 \\
0 & 0 & 2 & 0 & 0 \\
0 & 0 & 0 & 2 & 0 \\
0 & 0 & 0 & 0 & 2
\end{pmatrix} \, .
\end{equation}

The Laplacian matrix for $C_5$ is given by
\begin{equation}
\Delta(C_5)=
\begin{pmatrix}
2  & -1 & 0  &0  &-1 \\
-1 & 2  & -1 &0  &0 \\
0  & -1 & 2  &-1 &0 \\
0  & 0  & -1 &2  &-1 \\
-1 & 0  &0   &-1 &2
\end{pmatrix} \, .
\label{cm}
\end{equation}
Up to the dimensionful coefficient $f^2$,
this matrix is identified with 
the gauge boson $(mass)^2$ matrix (\ref{gauge-boson_mass-matrix}).
The structure of the theory can be generalized to the one
based on other complicated graphs.

Indeed, any theory space can be associated with a graph
and the $(mass)^2$ matrix for a field on a graph can be expressed by the graph
Laplacian
through the Green's theorem for a graph
\begin{equation}
(d\phi_1, d\phi_2)=(\phi_1, \Delta \phi_2)\,,
\end{equation}
where $\phi$ is a function that assigns a real value $\phi(v)$ to each vertex $v$ of
$G$ and
$(\phi_1, \phi_2)$ denotes the inner product $\sum_{v}\phi_1(v)\phi_2(v)$. The
difference defined on each edge
$e$ is
\begin{equation}
(d\phi)(e)\equiv \phi(t(e))-\phi(o(e))\,,
\end{equation}
where $t(e)$ is the terminus of $e$  and $o(e)$ is the origin of $e$.
$(d\phi_1, d\phi_2)$ is defined as $\sum_{e} d\phi_1(e)d\phi_2(e)$.
For example, a mass term of scalar fields can be constructed as
$f^2(d\phi, d\phi)=f^2(\phi, \Delta \phi)$.

The mass term for fermion fields can be written by using the
incidence matrix
$E$. 
For example, the lagrangian density of fermion fields can be written as\cite{KSJMP}
\begin{equation}
-(\bar{\psi}_R,\mbox{\it D\!\!\!\!/~}\psi_R)-(\bar{\psi}_L,\mbox{\it
D\!\!\!\!/~}\psi_L)-f[(\bar{\psi}_L,E^T\psi_R)+h. c.]\,,
\label{fermionlag}
\end{equation}
where the subscripts $L$ and $R$ stand for left-handed and right-handed fermions,
respectively.
Namely, the left-handed fermions are assigned to the edges while the right-handed
ones are assigned to the vertices. The $(mass)^2$ matrix for $\psi_R$ is expressed as
$f^2EE^T=f^2\Delta$ while that for $\psi_L$ is $f^2E^TE\equiv f^2\tilde\Delta$. The
matrix $\Delta$ and
$\tilde\Delta$ have the same spectrum up to zero modes. Thus the mass spectrum of
fermions governed by the lagrangian (\ref{fermionlag}) is also given by the eigenvalues
of the graph Laplacian (\ref{EE}). For the details, see Ref
\citen{KSJMP}.
                          
\section{Our previous work}\label{opw}
We have constructed models of one-loop finite induced gravity by using several
graphs in Ref \citen{KSPTP}.
With the help of knowledge in {spectral graph theory},
we can easily find that the UV divergent terms are concerned with the graph Laplacian
in DD or the theory on a graph.

We have observed important relations: The quadratic
divergence is proportional to  the trace of ($(mass)^2$  matrix)  and
the logarithmic  divergence is proportional to the trace of  ($[(mass)^2 matrix]^2$)
matrix.
For cancellation of the quartic divergence, we choose the particle content so that the
bosonic and fermionic degrees of freedom should be same. This
is the same consequence required from supersymmetry.

Therefore, the UV divergences can be controlled by the graph Laplacian
and we can construct the models of one-loop finite induced gravity from graph.
In the model of Ref \citen{KSPTP}, the one-loop finite Newton's constant is induced
and the positive-definite cosmological constant can also be obtained.

In flat-space limit, the one-loop vacuum energy has been calculated
for field theory on $C_N$.\cite{KSPTP} The finite part from various spin fields are:
\begin{eqnarray}
& &\mbox{\rm For a scalar field,}\qquad
V_0=-\frac{3}{4\pi^2}\left(\frac{f}{N}\right)^4Z(5,N)\,,\\
& &\mbox{\rm For a vector field,}\qquad
V_0=-\frac{9}{4\pi^2}\left(\frac{f}{N}\right)^4Z(5,N)\,,\\
& &\mbox{\rm For a Dirac field,}\qquad
V_0=+\frac{3}{\pi^2}\left(\frac{f}{N}\right)^4Z(5,N)\,,
\end{eqnarray}
where `a' field means that a lagrangian for a set of four-dimensional fields on
a graph. The function $Z(5,N)$ is defined as
\begin{equation}
Z(5,N)\equiv\sum_{q=1}^{\infty}\frac{1}{q\left(q^2-\frac{1}{N^2}\right)
\left(q^2-\frac{4}{N^2}\right)}\,.
\end{equation}

The inverse of the Newton constant G has also been computed for theory on
$C_N$.\cite{KSPTP} The finite part from various spin fields are:
\begin{eqnarray}
& &\mbox{\rm For a scalar field,}\qquad
(16\pi {\rm G})^{-1}=+\frac{1}{48\pi^2}\left(\frac{f}{N}\right)^2Z(3,N)\,,\\
& &\mbox{\rm For a vector field,}\qquad
(16\pi {\rm G})^{-1}=-\frac{1}{16\pi^2}\left(\frac{f}{N}\right)^2Z(3,N)\,,\\
& &\mbox{\rm For a Dirac field,}\qquad
(16\pi {\rm G})^{-1}=+\frac{1}{24\pi^2}\left(\frac{f}{N}\right)^2Z(3,N)\,,
\end{eqnarray}
where $Z(3,N)$ is defined as
\begin{equation}
Z(3,N)\equiv\sum_{q=1}^{\infty}\frac{1}{q\left(q^2-\frac{1}{N^2}\right)}\,.
\end{equation}
Note that $Z(5,\infty)=\zeta_R(5)$ and $Z(3,\infty)=\zeta_R(3)$, where $\zeta_R(z)$ is
Riemann's zeta function.

In the next section, we study self-consistent cosmological solutions for an
Einstein universe in the graph-based induced gravity model. 

\section{Self-consistent Einstein Universe} 
The static homogeneous, closed space is often called as an Einstein
universe. 
The solution for the static hot universe of positively curved space can be deduced from
Einstein equations with a cosmological constant and homogeneous matter in thermal
equilibrium  at finite temperature.

The back reaction problem was studied for various spin field in such a static curved
space at finite temperature.  Einstein equations were solved self-consistently in
Refs \citen{EU1,EU2}.
Self-consistent solutions were also investigated in higher-dimensional theory with
compactification\cite{KKEU}. 
The present work may be similar to Ref~\citen{KKEU}, because 
we should consider the excited states, which come from selected fields.

The metric of the spherically-symmetric, static homogeneous universe\cite{EU1,EU2,KKEU}
is given by 
\begin{equation}
ds^2=-dt^2+a^2[d\chi^2+\sin^2\chi(d\theta^2+\sin^2\theta d\phi^2)]\,,
\end{equation}
where 
$a$ is the radius of the spatial part of the universe $S^3$
and $0 \le \chi \le \pi$, $0 \le \theta \le \pi$ and $0 \le \phi < 2\pi$.
The scalar curvature of this spacetime is given by $R=6/a^2$.

In the static spacetime, it is known that the effective action can be interpreted as
the total free energy of the quantum fields.\cite{KKEU}
The one-loop part of the effective potential considered here is
given by the free energy in the canonical ensemble.

To evaluate the effective action at the one-loop level, we use the heat kernel
method\cite{hk}.
The free energy of a system of bosonic fields can be computed as the integration
over the parameter $t$, 
{\small\begin{equation}
F_b=-\frac{1}{2\beta}\int_{1/\Lambda^2}^\infty
\frac{dt}{t}\,\sum_{n=-\infty}^{\infty}\exp\left[-\left(\frac{2\pi
n}{\beta}\right)^2t\right]\left\{ {\rm
tr}\,\exp\left[-(-\nabla^2)t\right]\right\}\sum_i\exp\left[-M^2_i t\right]\,,
\label{fb}
\end{equation}}%
while the free energy of a system of fermionic fields can be computed as
{\small\begin{equation}
F_f=\frac{1}{2\beta}\int_{1/\Lambda^2}^\infty
\frac{dt}{t}\,\sum_{n=-\infty}^{\infty}\exp\left[-\left(\frac{2\pi
}{\beta}\right)^2\left(n+\frac{1}{2}\right)^2t\right] \left\{{\rm
tr}\,\exp\left[-(-\nabla^2)t\right]\right\}\sum_i\exp\left[-M^2_i t\right]\,.
\label{ff}
\end{equation}}%
Here $\beta$ is the inverse of the temperature of the system, that is $\beta=1/T$.
$\Lambda$ is the mass scale of the UV cutoff.
$M_p$ is the mass and $\nabla^2$ is the Laplacian on the $S^3$ (with radius $a$) of
the correspondent field.\footnote{Precisely speaking, the ``Laplacian'' is defined so
that the absolute value of its eigenvalue is the square of the eigen-frequency of
the corresponding field.}

The trace part $\{{\rm tr}\,\exp\left[-(-\nabla^2)t\right]\}$ can be evaluated by use
of the eigenvalues of the Laplacian. Thus we need the eigenvalues of the Laplacian
($\nabla^2$) on
$S^3$ with radius $a$, for various spin fields.\cite{EU1}

For a minimally-coupled scalar, the eigenvalues of $-\nabla^2$ are
\begin{equation}
\frac{\ell(\ell+2)}{a^2}\,,
\end{equation}
and the degeneracy of each eigenvalue is $(\ell+1)^2$, where $\ell=0,1,2,3,\ldots$.

For a vector field, 
two parts of spectra is known.
One is the transverse mode.
Its eigenvalues are
\begin{equation}
\frac{\ell^2}{a^2}\,,
\end{equation}
with the degeneracy $2(\ell^2-1)$, where $\ell=2,3,\ldots$.
Another is the longitudinal mode whose eigenvalues are
\begin{equation}
\frac{\ell(\ell+2)}{a^2}\,,
\end{equation}
with the degeneracy $(\ell+1)^2$, where $\ell=1,2,3,\ldots$.
Note that this spectrum is the same as that of a minimally-coupled scalar,
except for its zero mode.

The Laplacian eigenvalues for a Dirac fermion field,
where the Laplacian is interpretted as the square of the Dirac operator, are 
\begin{equation}
\frac{(\ell+1/2)^2}{a^2}\,,
\end{equation}
and the degeneracy is $4\ell(\ell+1)$, where $\ell=1,2,3,\ldots$.

The trace part $\{{\rm tr}\,\exp\left[-(-\nabla^2)t\right]\}$ for each field can be
evaluate as follows and has an asymptotic form for small $t$:\footnote{The
useful summation formulas are collected in Appendix.} 
{\small
\begin{equation}
\sum_{\ell=0}^\infty (\ell+1)^2\exp\left[-\frac{\ell(\ell+2)}{a^2}\,t\right]
=\frac{2\pi^2a^3}{(4\pi t)^{3/2}}\left(1+\frac{1}{a^2}\,t+\cdots\right)\,,
~~\mbox{for a scalar}\,,
\end{equation}
\begin{equation}
\sum_{\ell=2}^\infty 2(\ell^2-1)\exp\left[-\frac{\ell^2}{a^2}\,t\right]
=2\frac{2\pi^2a^3}{(4\pi t)^{3/2}}\left(1-2\frac{1}{a^2}\,t+\cdots\right)\,,
~~\mbox{for a transverse vector}\,,
\end{equation}
\begin{equation}
\sum_{\ell=1}^\infty (\ell+1)^2\exp\left[-\frac{\ell(\ell+2)}{a^2}\,t\right]
=\frac{2\pi^2a^3}{(4\pi t)^{3/2}}\left(1+\frac{1}{a^2}\,t+\cdots\right)\,,
~~\mbox{for a longitudinal vector}\,,
\end{equation}
\begin{equation}
\sum_{\ell=1}^\infty 4\ell(\ell+1)\exp\left[-\frac{(\ell+1/2)^2}{a^2}\,t\right]
=4\frac{2\pi^2a^3}{(4\pi t)^{3/2}}\left(1-\frac{1}{2a^2}\,t+\cdots\right)\,,
~~\mbox{for a Dirac spinor}\,.
\end{equation}
}

Now one finds that the trace part $\{{\rm tr}\,\exp\left[-(-\nabla^2)t\right]\}$
for each field
behaves as $\sim 1/t^{3/2}$ for $t\rightarrow 0$, while
the part expressed by the sum over $n$ in (\ref{fb}) or (\ref{ff}) behaves as
behaves as $\sim 1/t^{1/2}$ for $t\rightarrow 0$. Therefore, in generic cases, we
recognized that the free energy diverges as $\Lambda^4$ for a large cutoff $\Lambda$.

If the mass spectrum of a specific field is determined by a graph,
as briefly explained in \S 2 and in Refs \citen{KSPTP,KSJMP},
the last piece in (\ref{fb}) or (\ref{ff}) becomes
\begin{equation}
\sum_p\exp\left[-M^2_p t\right]={\rm Tr}\exp\left[-f^2\Delta(G) \,t\right]\,,
\label{mk}
\end{equation}
where $f$ is a mass scale and $\Delta(G)$ is the graph Laplacian for the graph $G$.
We found that, for a small $t$, the coefficients of $t^0$, $t^1$, and $t^2$ of
(\ref{mk}) depend on the number of vertices and edges and the degrees of
vertices.\cite{KSPTP,KSJMP}
Thus without enormous effort,\cite{IG} we can choose a particle content and graphs for
the mass spectrum of each field to cancel the $\Lambda^4$ and $\Lambda^2$ terms in the
total free energy of bosonic and fermionic fields.

As far as we consider only minimally-coupled fields to gravity,
we found a unique set of field contents for a finite free energy:
one minimally-coupled neutral scalar field, one vector field, and one Dirac field.
Here, `one' field means that there is a theory of the four-dimensional fields on a
graph.  Notice that this ratio is the same as the content of five-dimensional
supersymmetric theory including a vector multiplet.\cite{bk}
This particle (field) content is a condition of necessity for finite theory.
We select a graph whose Laplacian gives the mass spectrum for each field
to cancel the cutoff dependent parts in total.
We adopt a graph $G_S$ for a scalar field, $G_V$ for a vector, and $G_D$ for a Dirac
field, respectively, as a graph on which each field theory is based. 

\vspace{1mm}
We find a sufficient condition for the cancellation of UV divergences on graphs:
{\small 
\begin{equation}
{\rm Tr}\,D(G_S)={\rm Tr}\,D(G_V)={\rm Tr}\,D(G_D)~
{\rm and}~
{\rm Tr}\,D^2(G_S)={\rm Tr}\,D^2(G_V)={\rm Tr}\,D^2(G_D)\,,
\end{equation}
}%
for the above-mentioned field content.\cite{KSPTP}


Unfortunately, the existence of scalar zero mode leads to the divergence in the
integration at large $t$. This IR-like divergence can be regularized by the small
mass scale cutoff and it can easily seen from the form of (\ref{fb}) that only the
constant contribution ({\it i.e.}, independent of $\beta$ and $a$) to $\beta F_b$ is
small-mass-scale cutoff dependent. 

It is known that the quickest way to obtain the self-consistency equations is found
by using the total free energy $F$.\cite{EU1}
The self-consistent equation can be derived by  
\begin{equation}
\frac{\partial(\beta F)}{\partial \beta}=0\quad
{\rm and}\quad
\frac{\partial(\beta F)}{\partial a}=0\,,
\label{Einstein-equation}
\end{equation}
where the first equation corresponds to the $00$-component of the Einstein equation
with one-loop corrections
and the second corresponds to the diagonal component in a spatial direction.
Thus the extremal point of $\beta F(a,\beta)$ gives a solution to the self-consistent
equation and the solution is independent of the small-mass-scale cutoff mentioned in
the above paragraph.

Now we will consider specific models and will show the self-consistent
solutions.

We exhibit two models.
In our models, four-dimensional fields are defined on graphs consisting of one or
several cycle graphs. Here $C_N$ denotes a cycle graph
with $N$ vertices, equivalent to an $N$-sided polygon. Thus each degree of vertex is
two. Since we omit showing each lagrangian explicitly, please consult on our
papers\cite{KSPTP,KSJMP} for the detail.
We mention here only on the mass spectra are given by 
$f^2 \Delta(G_S)$, $f^2 \Delta(G_V)$ and $f^2 \Delta(G_D)$ as the $(mass)^2$ matrices
for minimal scalar fields, vector fields and Dirac fields, respectively.

We take the following notation: $G_a=G_1+G_2$ reads
$G_a=G_1\cup G_2$, $G_b=2 G_c$ reads $G_b=G_c\cup G_c$, and so on.
In these examples, graphs $G_a$ and $G_b$ are not simple graphs because of their
disconnected structure, in general.

The model~1 is now described as follows.
We consider that 
scalar fields are on $G_S=8 C_{N/2}$, vector fields on $G_V=4 C_N$ and Dirac fermions
on
$G_D=2 C_{N/2}+3 C_N$. Each graph has an equal number of vertices and edges, $4 N$. 
We find also ${\rm Tr}\,D(G_S)={\rm Tr}\,D(G_V)={\rm Tr}\,D(G_D)=8N$ and 
${\rm Tr}\,D^2(G_S)={\rm Tr}\,D^2(G_V)={\rm Tr}\,D^2(G_D)=16 N$.

On the other hand the model~2 is described as follows.
We consider that  
scalar fields are on $16 C_{N/4}+2 C_{N/2}$, vector fields on $5 C_N$ and Dirac
fermions on $4 C_{N/4}+3 C_N+2 C_{N/2}$.
Each graph has an equal number of vertices and edges, $5 N$.
We find also ${\rm Tr}\,D(G_S)={\rm Tr}\,D(G_V)={\rm Tr}\,D(G_D)=10 N$ and 
${\rm Tr}\,D^2(G_S)={\rm Tr}\,D^2(G_V)={\rm Tr}\,D^2(G_D)=20 N$.

We emphasize again that in each model,
Newton's constant and the cosmological constant are calculable 
in the criterion of Ref \citen{KSPTP} and are not given by hand as in the works of
Refs \citen{EU1,EU2}.

For the field on $C_N$, the trace of the part of kernel of the mass spectrum is given
by
\begin{equation}
\sum_p \exp(-M^2_p t)=N e^{-2f^2t}\sum_{q=-\infty}^{\infty}I_{qN}(2f^2t)\,,
\end{equation}
where $f$ is a mass scale in the model.

For $N \gg 1$ and $2f^2t \gg 1$, we find
\begin{equation}
\sum_p \exp(-M^2_p t)\sim \frac{N}{f}
\frac{1}{\sqrt{4\pi t}}\sum_{q=-\infty}^{\infty}\exp\left(-\frac{N^2q^2}{4f^2
t}\right)\,.
\end{equation}
This asymptotic formula holds as much like in the case for the KK spectrum on $S^1$.
Because the cancellation of divergence means that the region of small $t$ is
ineffective, the calculation of the free energy can be approximated as the one in KK
theories. The results given in the next section are on the model~1 and 2 in the limit
of
$N\rightarrow\infty$.%
\footnote{It is safe to say that the `approximation' of the large $N$ limit is valid
even for $N\approx 10$.\cite{KSepj}}
Thus the present calculation is much akin to the one in the case of KK
theories.\cite{KKEU}

We selected the content of each model so that the induced Newton constant becomes
positive. In the model 1 the induced vacuum energy vanishes in flat-space limit,
 while in the model 2 that is positive in the limit. Referring to the result of
Ref~\citen{KSPTP}, noted in \S~\ref{opw}, one can evaluate the one-loop vacuum energy
$V_0$ and the inverse of the Newton constant G in flat-space limit and for large $N$ as
follows:
\begin{eqnarray}
& &\mbox{Model 1:}\quad V_0=0\,, \quad(16\pi
{\rm G})^{-1}=\frac{7}{8\pi^2}\left(\frac{f}{N}\right)^2\zeta_R(3)\,,\\ 
& &\mbox{Model 2:}\quad 
V_0=\frac{279}{4\pi^2}\left(\frac{f}{N}\right)^4\zeta_R(5)\,,
\quad(16\pi {\rm G})^{-1}=\frac{133}{16\pi^2}\left(\frac{f}{N}\right)^2\zeta_R(3)\,.
\end{eqnarray}
The feature that $V_0$ and G is merely non-negative in our toy models is essential
and we will not go into persuiting models with realistic values for parameters here.

\section{Results}
\begin{figure}[h]
\centering
\includegraphics
{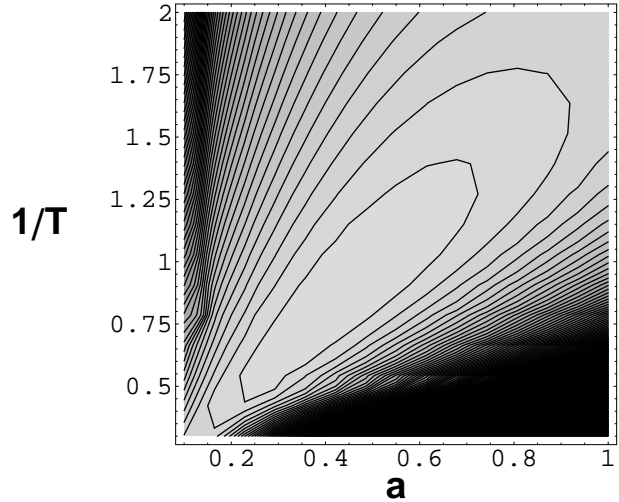}
\caption{%
A contour plot of $\beta F$ in the first model,
in which scalar on $8 C_{N/2}$, vectors on $4 C_N$, and 
Dirac fermions on $2 C_{N/2}+3 C_N$.
A solution of the self-consistent equation can be found at the maximum point.
}
\label{fig2}
\end{figure}
\begin{figure}[h]
\centering
\includegraphics
{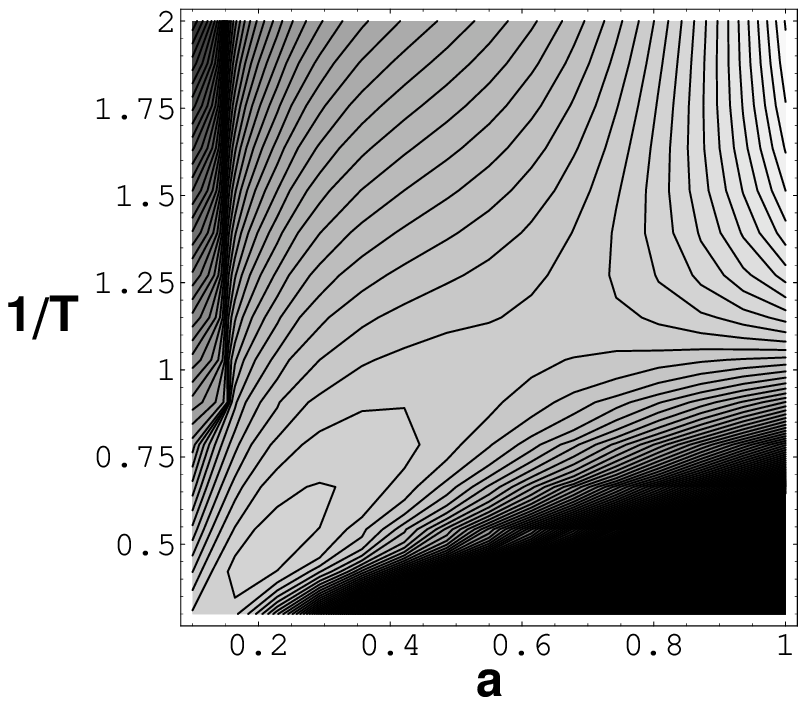}
\caption{%
A contour plot of $\beta F$ in the second model, in which
scalars on $16 C_{N/4}+2 C_{N/2}$, vectors on $5 C_N$ and Dirac fermions on 
$4 C_{N/4}+3 C_N+2 C_{N/2}$.
Two solutions of the self-consistent equation can be found at the maximum and at the
saddle point.
}
\label{fig3}
\end{figure}

We show the contour plots for $\beta F$ obtained by numerical calculations, whose
extrema give self-consistent solutions. The horizontal axis indicates the scale factor
$a$, while the vertical one indicates the inverse of temperature $\beta=1/T$.
The scale of each
axis is in the unit of $N/f$.

There are two regimes introduced in Ref \citen{EU1}. Corresponding to large $Ta$, we
have the Planck regime, and for small $Ta$ we have the Casimir regime.
They are characterized by the leading contributions $F_l$ in the total free energy,
$F_l\propto -a^3T^4$ in the Planck regime and $F_l\propto -a^3\times 1/a^4$ in the
Casimir regime.

At high temperature, 
the energy of the black-body radiation in the Planck distribution
overcomes in the one-loop contribution from quantum fluctuations.
At low temperature, 
finite-temperature piece of quantum correction is negligible and the
Casimir density dependent on the scale factor
$a$ as well as zero-temperature induced gravity is left.

We exhibit $\beta F$
for the first model in Fig \ref{fig2}
and for the second in Fig \ref{fig3}, for large $N$.

In the first model, the induced cosmological constant,
which is independent of $a$ or $\beta$, 
vanishes and the solution 
can be found at the maximum of $\beta F$, corresponding to be
in the Casimir regime.\cite{EU1}

In the second model,
the solution in the Casimir regime and the solution in the Planck regime are both
found.

In both models, no solution corresponding the minimum of $\beta F$ can be found.
Because the instability of hot closed universe is well known, this result can be
expected. The result tells us that the one-loop quantum effect cannot stabilize the
solution.

\section{Summary and Outlook}
We have studied self-consistent Einstein universe in induced gravity models
constructed based on the `graph theory space'. The solution can be systematically
obtained by the knowledge of the graph structure.
We find that the various pattern of regimes can been seen according to the model,
{\it i.e.}, the choice of graphs.

As the future works,
we should discuss the possibility of obtaining the small cosmological constant and
the large Planck scale
in a model that scalar fields are on $4 G_{(1)}$, vector fields on $4 G_{(2)}$ and
Dirac fermion on $G_{(1)}+3 G_{(2)}$,
where the number of vertices in $G_{(1)}$ and that of $G_{(2)}$ are equal%
.
We also should investigate the model with the time-dependent scale factor, $a(t)$.

In the present analysis, we have constructed models by using cycle graphs, 
but we are also interested in the model of general graphs.
For a $k$-regular graph,
trace formula\cite{tf} is useful 
if we have a single mass scale.
Field theory on weighted graphs, which corresponds to warped spaces
in the continuous limit or not, is worth studying.

Turning to looking at the properties of matter fields at finite temperature,
to study the way how the condensation or degenerate matters in a closed space\cite{cd}
affects the mechanism of induced gravity is also an interesting subject.

There are other possible exotic ideas.
A quasi-continuous mass spectrum is conceivable
and
dynamics of graphs such as Hosotani mechanism\cite{Hosotani} is also reasonable
to think. A discrete `flux'-like object may be associated to a closed circuit in a
graph.
 
Finally, we expect that the knowledge of spectral graph theory produces useful results
on deconstructed theories in all directions
and open up another possibilities of gravity models.


\section*{Acknowledgements}
We would like to thank K.~Kobayashi for discussion.

\appendix
\section{
formulas on summations
} 

\begin{eqnarray}
\sum_{n=-\infty}^\infty \exp\left[-\left(\frac{2\pi}{\beta}\right)^2 n^2\, t\right]
&=&\frac{\beta}{\sqrt{4\pi t}}
\sum_{n=-\infty}^{\infty}\exp\left(-\frac{\beta^2n^2}{4t}\right)\,.\\
\sum_{n=-\infty}^\infty \exp\left[-\left(\frac{2\pi}{\beta}\right)^2 
\left(n+\frac{1}{2}\right)^2t\right]
&=&\frac{\beta}{\sqrt{4\pi t}}
\sum_{n=-\infty}^{\infty}(-1)^n \exp\left(-\frac{\beta^2n^2}{4t}\right)\,.
\end{eqnarray}

\begin{equation}
\sum_{\ell=0}^\infty (\ell+1)^2\exp\left[-\frac{\ell(\ell+2)}{a^2}\,t\right]
=\frac{2\pi^2a^3}{(4\pi t)^{3/2}}e^{t/a^2}
\sum_{\ell=-\infty}^{\infty}
\left(1-\frac{2\pi^2a^2\ell^2}{t}\right)e^{-\frac{\pi^2a^2\ell^2}{t}}\,.
\end{equation}
\begin{equation}
\sum_{\ell=2}^\infty 2(\ell^2-1)\exp\left[-\frac{\ell^2}{a^2}\,t\right]
=2\frac{2\pi^2a^3}{(4\pi t)^{3/2}}
\sum_{\ell=-\infty}^{\infty}
\left(1-\frac{2t}{a^2}-\frac{2\pi^2a^2\ell^2}{t}\right)
e^{-\frac{\pi^2a^2\ell^2}{t}}+1\,.
\end{equation}
\begin{equation}
\sum_{\ell=1}^\infty 4\ell(\ell+1)\exp\left[-\frac{(\ell+1/2)^2}{a^2}\,t\right]
=4\frac{2\pi^2a^3}{(4\pi t)^{3/2}}\sum_{\ell=-\infty}^{\infty}(-1)^\ell
\left(1-\frac{t}{2a^2}-\frac{2\pi^2a^2\ell^2}{t}\right)
e^{-\frac{\pi^2a^2\ell^2}{t}}\,.
\end{equation}


\end{document}